\documentclass[preprint]{aastex}
\usepackage[]{natbib}
\usepackage[]{verbatim}

\newcommand{\feh}{\ensuremath{\rm [Fe/H]}}

\newcommand{\masyr}{mas\,yr$^{-1}$}
\newcommand{\kms}{km\ s$^{-1}$}

\newcommand{\msun}{M$_{\odot}$}

\newcommand{\vc}{V474 Car}
\newcommand{\vsini}{$v$\,sin\,$i$}

\slugcomment{accepted to Astronomical Journal}
\shorttitle{V474 Car}
\shortauthors{Bubar et al.}
\begin{document}

\title{V474 Car: A Rare Halo RS CVn Binary in Retrograde Galactic Orbit}

\author{Eric J. Bubar, Eric E. Mamajek}

\affil{University of Rochester, Department of Physics \& Astronomy,
  Rochester, NY, 14627-0171, USA}

\author{Eric L. N. Jensen}

\affil{Swarthmore College, Department of Physics \& Astronomy, 500
  College Avenue, Swarthmore, PA 19081, USA}

\author{Frederick M. Walter}

\affil{Department of Physics \& Astronomy, Stony Brook University,
  Stony Brook, NY 11794-3800}

\begin{abstract}
  We report the discovery that the star V474 Car is an 
  extremely active, high velocity halo RS CVn system. The star was 
  originally identified as a
  possible pre-main sequence star in Carina, given its enhanced
  stellar activity, rapid rotation (10.3 days), enhanced Li, and
  absolute magnitude that places it above the main sequence. However,
  its extreme radial velocity (264 km s$^{-1}$) suggested that this
  system was unlike any previously known pre-MS system. Our detailed
  spectroscopic analysis of echelle spectra taken with the CTIO 4-m
  finds that V474 Car is both a spectroscopic binary with orbital
  period similar to the photometric rotation period, and metal
  poor ([Fe/H] $\simeq -$0.99). The star's Galactic orbit is extremely
  eccentric (e $\simeq$ 0.93) with perigalacticon of only $\sim$0.3
  kpc of the Galactic center - and its eccentricity and smallness of
  its perigalacticon are only surpassed by $\sim$0.05\%\, of local F/G-type
  field stars. The observed characteristics are consistent with V474
  Car being a high velocity, metal poor, tidally-locked
  chromospherically active binary (CAB), i.e.\ a halo RS CVn binary,
  and one of only a few such specimens known.
\end{abstract}


\section{Introduction}
\label{section:intro}

In the course of a search for nearby pre--main-sequence stars among
Hipparcos catalog stars with strong X-ray emission \citep{Jensen10},
E.~J. encountered the unusual star V474 Car (HIP~44216, CPD-62 1150) 
which spectroscopically appeared to be remarkably
similar to a G-type T Tauri star, with saturated X-ray emission and
some H$\alpha$ emission, but with a remarkably large radial
velocity ($v_r$ $\simeq$ 250 \kms). As there were indications that
this star might be a runaway T Tauri star, we conducted some follow-up
observations to further characterize it.

There has been very little previous study of V474 Car. The star was
listed as a high proper motion star ($\mu$ $\simeq$ 0\farcs2/yr)
multiple times by Luyten, and catalogued by \citet{Luyten79} as L
139-59, LTT 3338, and NLTT 20802. The star appeared in the 74th name
list of variable stars as V474 Car \citep{Kazarovets99} and is entry 
ASAS 090023-6300.1 in the All Sky Automated Survey (ASAS)
Catalog of Variable Stars \citep{Pojmanski05}.  The ASAS catalog
lists V474 Car as a BY Dra-type star with a period of 10.312 days,
V-band amplitude of 0.16 mag and maximum brightness of V = 9.94 mag.
\citet{Torres06} listed V474 Car as SACY 525 in their spectroscopic
survey of southern ROSAT X-ray sources, and classified the star as being a
G0Ve star with large radial velocity ($v_r$ = 247.8 \kms), slow
projected rotation, but with
lithium (EW(Li I $\lambda$6707.8) = 100 m\AA), and weak
H$\alpha$ emission (EW(H$\alpha$) = $-1.5$ \AA).  Hence, this star
simultaneously shows paradoxical characteristics of both youth and old
age.

Here we report our observations of the mysterious star V474 Car. We
conclude that it is an extremely active, tight binary with an
abundance pattern and space velocity consistent with belonging to the
Galactic halo population. The photometric period for V474 Car matches
the orbital period measured for the spectroscopic binary, confirming
that V474 Car is a tidally-locked, active binary, i.e.\ an RS CVn
system.

\begin{deluxetable}{lcl}
\setlength{\tabcolsep}{0.03in}
\tablewidth{0pt}
\tablecaption{Properties of V474 Car \label{tab:props}}
\tablehead{Property & Value & Ref.}
\startdata 
Parallax       & 8.07\,$\pm$\,1.07 mas       & 1\\
Distance       & 124$^{+19}_{-15}$ pc        & 1\\
RA(J2000)      & 09:00:23.24                 & 1\\
Dec(J2000)     & $-$63:00:04.3                 & 1\\  
$\mu_{\rm \alpha}$          &  114.54\,$\pm$\,0.93 \masyr & 1\\
$\mu_{\rm \delta}$         & $-$111.79\,$\pm$\,0.99 \masyr & 1\\
V              & 10.24 mag                   & 2\\
V$_{\rm max}$  &  9.94 mag                   & 4\\
V$_{\rm mean}$ & 10.03 mag                   & 4\\
J              & 8.184\,$\pm$\,0.023 mag     & 5\\
H              & 7.626\,$\pm$\,0.042 mag     & 5\\
K$_{\rm s}$    & 7.470\,$\pm$\,0.026 mag     & 5\\
B$-$V          & 0.904\,$\pm$\,0.015 mag     & 2\\
V$-$I$_{\rm C}$          & 0.95\,$\pm$\,0.02 mag       & 2\\
Period(phot)   & 10.30 days                  & 2\\
...            & 10.312 days                 & 4\\
Spec. Type     & G0Ve                        & 3\\
Soft X-ray flux & 0.195\,$\pm$\,0.025 ct/s   & 6\\
...             & 1.84\,$\pm$\,10$^{-12}$ erg/s/cm$^2$ & 6\\
HR1             & 0.21\,$\pm$\,0.12          & 6\\
HR2             & 0.23\,$\pm$\,0.15          & 6\\
L$_{\rm X}$     & 10$^{30.53}$ erg/s         & 6
\enddata
\tablecomments{References: (1) \citet{vanLeeuwen07}, distance is
  inverse of parallax, position is for epoch J2000 on ICRS, (2)
  \citet{Perryman97}, (3) \citet{Torres06}, (4) \citet{Pojmanski05},
  (5) \citet{Cutri03}, (6) \citet{Voges99}, conversion to X-ray flux
  and luminosity using \citet{Fleming95}, and luminosity using
  parallax from \citet{vanLeeuwen07}. All ROSAT X-ray fluxes are in
  the 0.1--2.4 keV band.}
\end{deluxetable}

\section{Observations \label{sec:obs}}

We took spectra of V474 Car with the 4-meter Blanco telescope at CTIO
on 2001 April 7 and 2003 April 14, using the echelle spectrograph. The
spectrograph covered a wavelength range from $4800 $ \AA\ to $8400 $
\AA\ with a measured spectral resolving power of $R \sim 40,000$
and typical S/N$\sim$100-130. The
spectra were reduced in IRAF\footnote{IRAF is distributed by the
National Optical Astronomy Observatories, which are operated by the
Association of Universities for Research in Astronomy, Inc., under
cooperative agreement with the National Science Foundation.}
according to standard procedures.  Observations of several radial
velocity standards on each night indicates an overall radial velocity
precision in high S/N spectra of better than 1 \kms.  

After noticing the large and variable radial velocity in the first two
observations, we also monitored \vc\ with the spectrographs on
the CTIO 1.5m telescope over a number of nights in April--June 2009
and January--February 2010.

We obtained 15 spectra with the echelle 
spectrograph\footnote{http://www.ctio.noao.edu/$\sim$atokovin/echelle/index.html}.
The echelle, formerly on the Blanco 4m, is fiber-fed.  The
detector is a 2k SITe CCD with 24 \micron\ pixels.  The full-chip
readout gives coverage from 4020 to 7300\AA\ at a resolving power of
about 25,000 to 30,000 for the 100 and 130 $\mu$m slits used.  Integration times 
were typically 1 hour, in three 20 minute integrations, except on one night when we only took
a single exposure.  Data reduction utilized a basic echelle reduction 
package\footnote{http://www.astro.sunysb.edu/fwalter/SMARTS/smarts\_15msched.html\#ECHpipeline}.

We also obtained six low dispersion spectra using the long-slit RC
spectrograph. These were used for spectral classification, to
determine the stellar activity level, and to search for evidence of a
cool companion.  We used a 1.5~arcsec slit and four standard low dispersion 
spectroscopic setups (26/Ia, 47/Ib, 47/II, 58/I).
We obtained three observations with integration times between 300 and 400
seconds, depending on the setup and the target brightness.  The three
observations were median-filtered to minimize contamination by cosmic
rays.  We reduced the data using our spectroscopic data reduction 
pipeline\footnote{http://www.astro.sunysb.edu/fwalter/SMARTS/smarts\_15msched.html\#RCpipeline}.

\section{Analysis \label{sec:anal}}

At first glance this star is puzzling, since it presents both
pre--main-sequence characteristics and halo population
characteristics.  If the star is placed on an HR diagram, assuming
that reddening corrections are negligible and that the secondary
contributes little light in the $V$ band, its position is consistent
with a pre--main-sequence classification (and indeed this is why we
initially selected it for observation).  In addition, we measure
modest H$\alpha$ emission (equivalent width$=-1.25$ {\AA}) and find a
lithium equivalent width from the $\lambda$6707.8 \ion{Li}{1} line of
91.1 m{\AA}.  All of these indicators are consistent with youth, which
could imply that the star may be a ``runaway'' post-T Tauri star.
However, the kinematics are also similar to that of halo stars.  In
order to clarify the status of this potentially interesting object, we
now present a variety of analyses.  We first discuss the kinematics,
including disentangling the effects of binary orbital motion from
those of overall space motion of the system.  We then discuss the
star's X-ray emission, and we conclude this section with a high
resolution spectroscopic abundance analysis.

\subsection{Binary Orbit}
\label{section:orbit}

Our first two observations, in 2001 and 2003, gave radial velocities
differing by 16 \kms, suggesting that the system is a spectroscopic
binary, so we made additional radial velocity observations.  The
measured velocities are given in Table \ref{table:velocities}.  We fit
the radial velocity data using the Binary Star Combined Solution
software \citep{Gudehus2001} and our own custom-written IDL code; both
gave the same solution.

\begin{figure}
\includegraphics[width=3.5in]{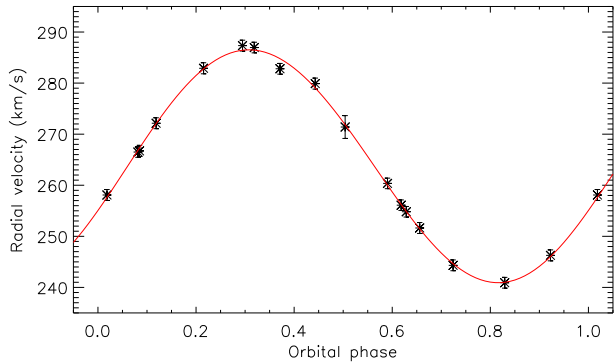}
\caption{Phased radial velocity measurements of V474 Car, with the
  best-fit binary orbital solution ($P=10.1944$ days; solid line) superimposed.
\label{figure:binary_orbit}}
\end{figure}

The best-fit phased radial velocity curve is shown in Figure
\ref{figure:binary_orbit}, and the orbital parameters are given in
Table \ref{table:orbital_params}.  The eccentricity is formally
consistent with a circular orbit; forcing a circular orbit in the fit
gives the same orbital parameters to within the uncertainties given in
Table \ref{table:orbital_params}.  In doing the fitting, we found that
the three data points taken in January and February of 2010 were
systematically low compared to the best-fit orbit.  The camera for the
spectrograph was removed from the dewar in December 2009, and we
suspect that our reduction pipeline does not yet fully correct for a
small rotation introduced when this change was made.  Thus, we have
applied an empirical $+3.9$ \kms\ correction to those three
datapoints, both in Table \ref{table:velocities} and in the final
orbital fit.  However, even if we do not apply that correction, the
orbital parameters are unchanged to within their quoted uncertainties.

\begin{figure}
\includegraphics[width=3.5in]{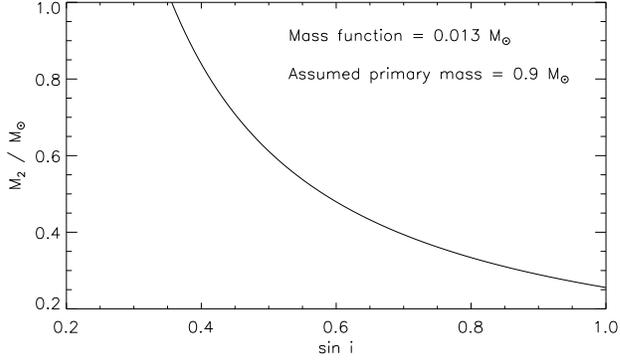}
\caption{Constraints on the mass of the secondary star, assuming a
  primary mass of 0.9 $M_\sun$.
\label{figure:secondary_mass}}
\end{figure}

The orbital fit yields a mass function of $0.013$ M$_\sun$, and thus
this can also be used to place limits on the secondary mass.  Based on our
derived stellar parameters for the primary (Sec.\
\ref{section:spec-analysis}), we interpolated Yale-Yonsei isochrones with a metallicity of
our derived value ([Fe/H]=-0.97: discussed below) to yield an
approximate primary mass of 0.9 $M_\sun$ \citep{Demarque04}.  As shown in Figure
\ref{figure:secondary_mass}, this constrains the secondary mass to
greater than 0.25 M$_\sun$.   
Thus, it is likely to be a late K or M
type companion, and it is plausible (especially for larger values of
$\sin i$) that we would not see any sign of it in our spectra.

We can place constraints on the inclination of the primary star's rotation
axis by comparing the inferred stellar rotation period with its projected
equatorial rotational velocity $v \sin i$ and also taking the star's radius
into account.  The Yale-Yonsei isochrones \citep{Demarque04} yield an
approximate primary stellar radius of 1.9 $R_\sun$, which is consistent
with a radius of 1.5-2.0 R$_\sun$ inferred from the Barnes-Evans
relation \citep{Barnes76} (uncertainty comes solely from parallax
and neglects systematic metallicity or interstellar reddening effects).  
Combining this with $P_{\rm rot}$
and $v \sin i$ (Table \ref{table:pp}) suggests a high value for $\sin i$ (within $1.7 \sigma$
of 1).  Assuming that the stellar rotation axis
is perpendicular to the binary orbit, this would imply a nearly edge-on orbit.
Though no eclipses are evident in the photometric data, an edge-on orbit would give a companion
mass of 0.26 $M_\sun$.  While Yale-Yonsei models do not go to this low 
of a mass, the lowest mass available in these models (0.40 $M_\sun$ yield a secondary stellar 
radius of 0.17 $R_\sun$.  As a plausibility study, this would imply
an eclipse depth of only 0.8 \% ,.which is undetectable given the existing
photometry.  It is worth noting however,
given our derived secondary characteristics, that it is possibly a white dwarf.

\begin{deluxetable}{ccc}
\tablewidth{0pt}
\tablecaption{Measured radial velocities \label{table:velocities}}
\tablehead{HJD & $v_{\rm r}$ (\kms) & $\sigma_{\rm v}$ (\kms) }
\startdata    
 2452006.59387  &  266.7  & 0.4 \\
 2452743.51096  &  282.8  & 0.3 \\
 2454958.60290  &  251.6  & 0.4 \\
 2454966.62641  &  279.9  & 0.4 \\
 2454969.49097  &  244.3  & 0.3 \\
 2454970.56424  &  240.9  & 0.3 \\
 2454971.51535  &  246.3  & 0.4 \\
 2454972.49168  &  258.1  & 0.4 \\
 2454973.51364  &  272.1  & 0.4 \\
 2454974.49883  &  282.9  & 0.4 \\
 2454985.50324  &  287.3  & 0.4 \\
 2454988.51663  &  260.4  & 0.2 \\
 2454993.52470  &  266.5  & 0.3 \\
 2455019.48989  &  254.8  & 0.3 \\
 2455201.71863  &  271.4\tablenotemark{a}  & 2.0 \\
 2455240.60836  &  287.0\tablenotemark{a}  & 0.3 \\
 2455243.66034  &  256.1\tablenotemark{a}  & 0.2 
\enddata
\tablenotetext{a}{Offset of $+3.9$ \kms\ added; see text.}
\end{deluxetable}

\begin{deluxetable}{lc}
\tablewidth{0pt}
\tablecaption{Spectroscopic Binary Solution \label{table:orbit}}
\tablehead{Property & Value}
\startdata    
Systemic radial velocity $\gamma$ & $ 263.8 \pm 0.3$ \kms \\
Velocity semi-amplitude $K_1$          & $22.8 \pm 0.5 $ \kms \\
$a_1 \sin i$   & $0.0214 \pm 0.0005$ AU  \\
$e$            & $0.014 \pm 0.018 $ \\
Period (days)  &  $10.1944 \pm 0.0003$ 
\enddata
\label{table:orbital_params}
\end{deluxetable}

\subsection{Kinematics}
\label{section:kinematics}

Having determined the system's center-of-mass radial velocity from the
binary orbital solution, we can now investigate its three-dimensional
kinematics more fully.

The proper motions for V474 Car from \citet{Perryman97},
\citet{Hog00}, \citet{vanLeeuwen07}, \citet{Roser08}, and
\citet{Zacharias10} are all within $\pm$2 \masyr\, of each other. To
estimate the Galactic Cartesian velocity for V474 Car, we combine the
position, proper motion, and parallax from \citet{vanLeeuwen07} with
the systemic radial velocity estimated in \S\ref{section:orbit}. The
resulting velocity is $U, V, W$ = (137, $-241$, $-41$) $\pm$ (10, 6,
3) \kms, where $U$ is pointed towards the Galactic center, $V$ is in
the direction of Galactic rotation ($\ell$, $b = 90\arcdeg$,
$0\arcdeg$), and $W$ is towards the North Galactic Pole ($b =
+90\arcdeg$).  Note that the $V$ velocity is almost directly opposite
the Galactic rotational velocity of $\sim 220$ \kms\ for disk stars in
the solar neighborhood, indicating that V474 Car has a retrograde
Galactic orbit with negligible rotational motion.

We used the \emph{orbit} packages within the {\it NEMO: Stellar
  Dynamics Toolbox} \citep{Teuben95} to integrate the orbit of this
star in a model Galactic potential (potential 1 taken from
\citealt{Dehnen98}).  From these orbit integrations we find an average
perigalacticon of 0.32 kpc and an apogalacticon of 9.20 kpc with an
eccentricity of 0.93 and an orbital period of 202 Myr. Hence, we are
catching V474 shortly after apogalacticon as it is ``falling'' towards
the Galactic center (where it should pass within $\sim2\fdg 3$
of Sgr A* during perigalacticon). 

\subsection{X-ray Emission}

V474 Car is detected as an X-ray source (1RXS J090022.8$-$630012) in
the ROSAT All-Sky Bright Source Catalog (RASS-BSC; \citealt{Voges99}).
While it is situated 9\arcsec\ away from the optical star, the RASS
X-ray source has an 8\arcsec\ positional error, consistent with V474
Car being the optical counterpart. V474 Car is the brightest optical
counterpart within 112\arcsec\ of the X-ray source, and as V474 Car is
clearly a chromospherically active, close binary, it is undoubtedly
responsible for the X-ray emission. The X-ray counterpart was imaged
over a total exposure time of 733 seconds on 14 June 1996, and has
measured X-ray properties listed in Table \ref{tab:props}. Using the
RASS-BSC count rate and HR1 hardness ratio, and the energy conversion
factor equation of \citet{Fleming95}, we estimate the observed soft
X-ray flux in the ROSAT passband (0.1-2.4 keV) of
1.84\,$\pm$\,10$^{-12}$ erg s$^{-1}$ cm$^{-2}$.  Combining the X-ray
flux with the revised Hipparcos parallax from \citet{vanLeeuwen07}, we
estimate the X-ray luminosity to be $L_{\rm X} = 10^{30.53}$ erg
s$^{-1}$. Through comparing the hardness ratios HR1 and HR2 for X-ray
emitting coronal plasma models from \citet{Neuhauser95}, V474 Car
appears to have coronal temperatures of $\sim$10--17 MK, with the
X-ray flux likely observed through negligible interstellar absorption
(N(H) $<$ 10$^{19}$ cm$^{-2}$). In comparing the hardness ratios HR1
and HR2 from \citet{Voges99} for a sample of nearby RS CVn systems
\citep{Dempsey93}, we find that typical RS CVns have median values of
HR1 = 0.03 ($\pm$0.12 ; 1$\sigma$) and HR2 = 0.16 ($\pm$0.09 ;
1$\sigma$).  Comparing the hardness ratios for V474 Car listed in
Table \ref{tab:props} with these mean values, we find that V474 Car's
X-ray colors are consistent (within the uncertainties) with those of
other RS CVn systems.  V474 Car's hardness ratios are at the high end
of the distribution, indicating relatively hot plasma.  This may be
due to the low metallicity of the system, which can inhibit efficient
cooling of the coronal plasma \citep{Fleming96}.

\subsection{Optical Variability}
\label{section:variability}

As noted in \S\ref{section:intro}, the All Sky Automated Survey (ASAS)
Catalog of Variable Stars \citep{Pojmanski05} show V474 Car to be
variable with period 10.312 days, V-band amplitude of 0.16 mag, and
maximum brightness of V = 9.94 mag.  This period is similar to the
binary orbital period, suggesting that the stars' rotation periods are
tidally synchronized with the binary orbit.  While the best-fit
photometric period of 10.3 days is slightly longer than the binary
orbital period of 10.2 days, it is still possible to reconcile the two
periods with synchronized rotation if the dominant spots are at
non-equatorial latitudes on the star and there is a modest amount of
differential rotation.

\begin{figure}
\includegraphics[width=3.5in]{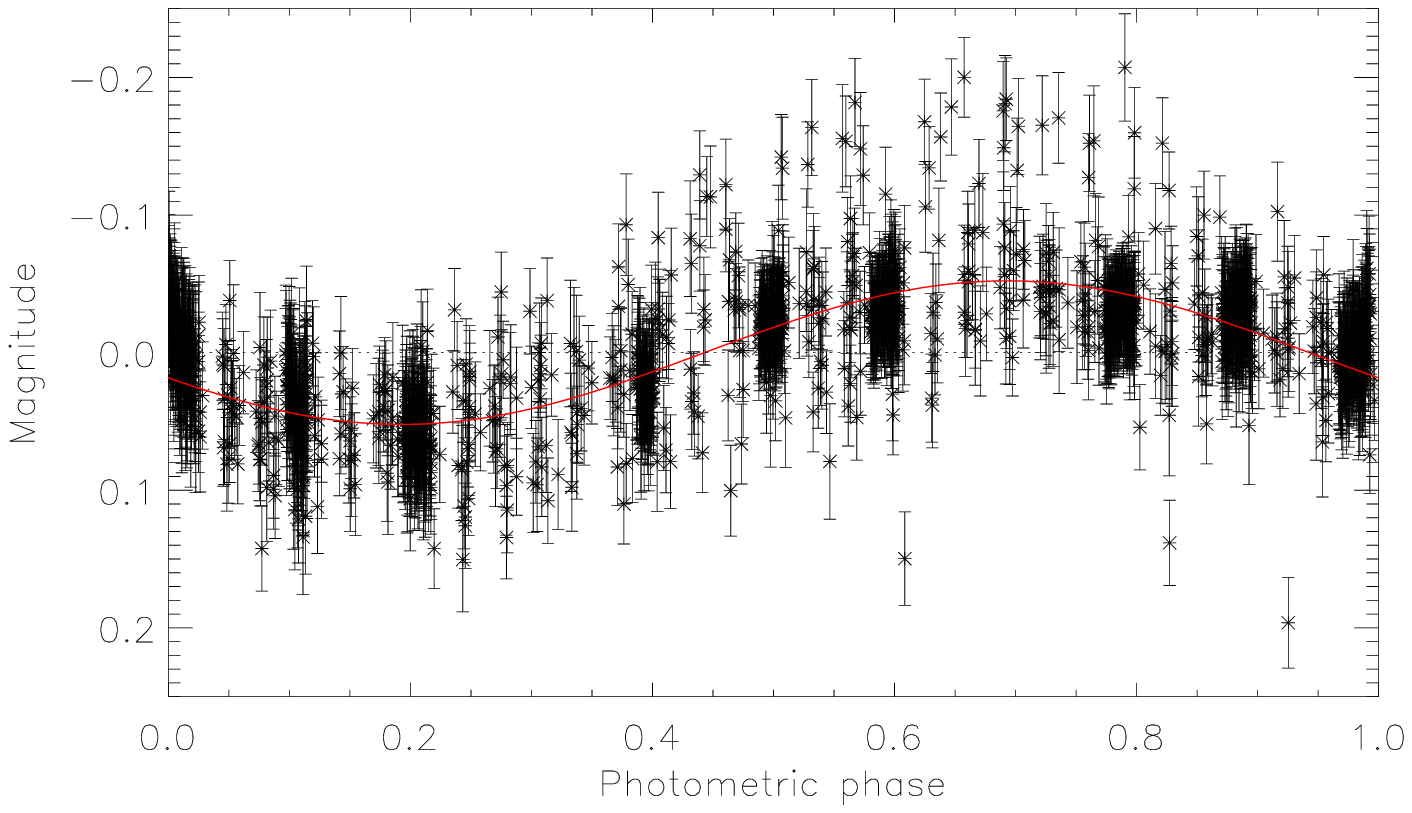}
\includegraphics[width=3.5in]{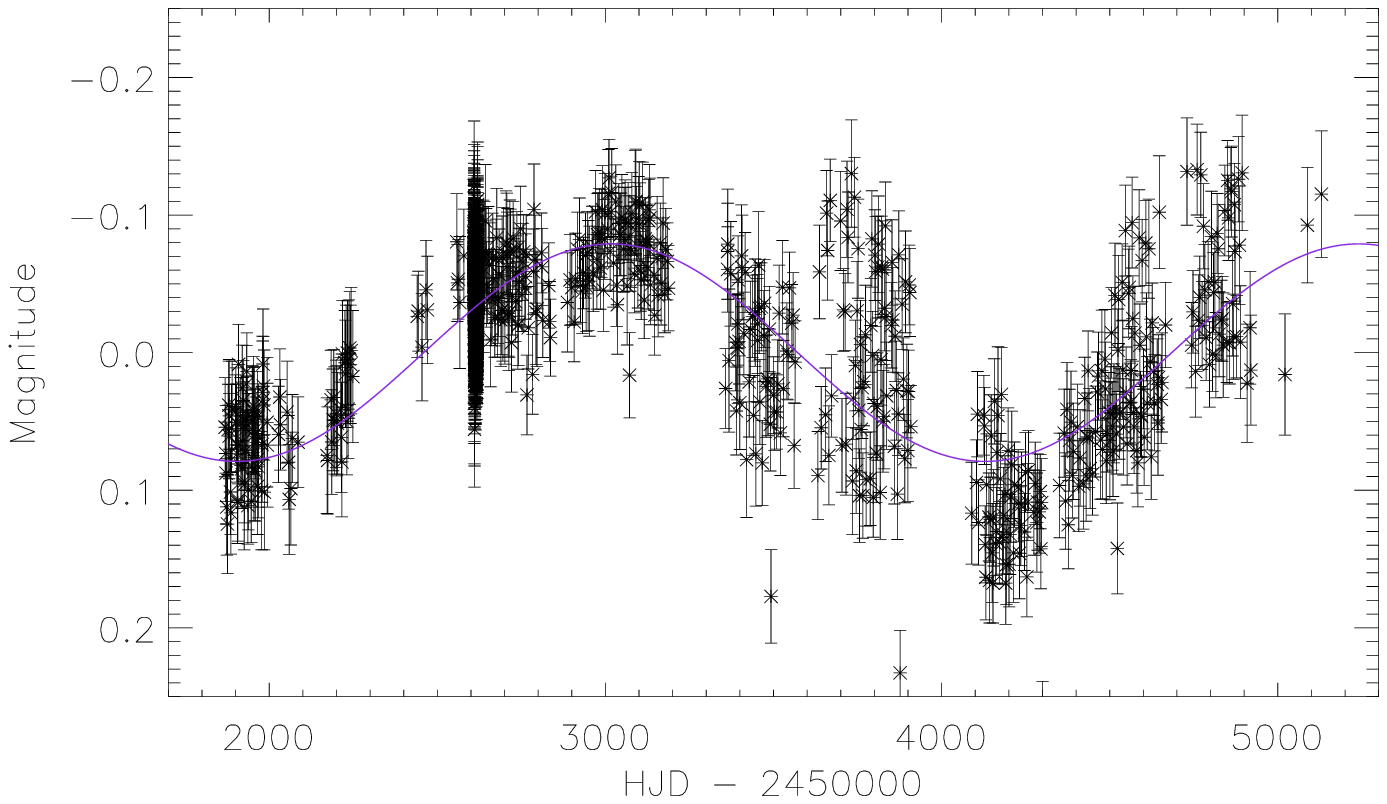}
\caption{ASAS V-band photometry of V474 Car, showing the periodic
  variations.  Top: Short-period variability, phased to $P=10.31$ days, similar to the
  binary orbital period.  Bottom: Long-term variability, suggesting a
  solar-like magnetic activity cycle of $\sim 6$ years.  Each plot
  shows the data with the other period removed.
  \label{figure:phot}}
\end{figure}

When the 10-day period is removed from the light curve, a much longer
term variation with an amplitude of $\sim 0.2$ mag remains (Figure
\ref{figure:phot}).  The modulation appears to be roughly sinusoidal
with a period of $\sim 6$ years.  Since the span of the data is only
about 9 years it is not possible to tell if the variation is truly
periodic.  However, both the inferred period and amplitude are
consistent with long-term activity cycles seen on other solar-type
stars \citep{Lockwood07}, suggesting that we may be observing the
analog of our Sun's 11-year magnetic activity cycle.

\begin{deluxetable}{ll}
\tablecaption{Derived Parameters for V474 Car\label{table:pp}}
\setlength{\tabcolsep}{0.03in}
\tablewidth{0pt}
\tablehead{Property & Value}
\startdata    
$M_{V}$                 & 4.56\\
Bolometric Correction   & -0.22 \citep{Flower96}\\
$M_{Bol}$               & 4.34\\
Mass$_{Primary}$        & 0.90 $\pm$ 0.02 $M_{\sun}$\\
Radius$_{Primary}$      & 1.90 $\pm$ 0.94 $R_{\sun}$\\
Luminosity$_{Primary}$  & 2.69 $\pm$ 0.77 $L_{\sun}$\\
T$_{\rm Excitation}$    & 5238 $\pm$ 152 K\\
$\log g$                & 3.64 $\pm$ 0.51\\
\feh\                  & $-0.99$ $\pm$ 0.16\\
Microturbulent Velocity & 2.52 $\pm$ 0.44 \kms\\
\vsini\                & 11.8 $\pm$ 3.0 \kms\\
EW(Li I $\lambda$6707)  & 91 m{\AA}\\
EW(H$\alpha$)           &$-1.25$ {\AA}
\enddata
\end{deluxetable}

\subsection{Spectroscopic Analysis}
\label{section:spec-analysis}

We utilized the echelle observations to derive a maximum projected rotational
velocity.  The \vsini\ was determined from a calibration of cross-correlation peak width 
versus \vsini\ for rotational velocity standards from \citet{Soderblom89} (HD 165185 :7.0 \kms, 
HD 206860: 10.2 \kms, HD 204121: 18.5 \kms, HD 134083: 45.0 \kms).  Macroturbulence
is included inasmuch as it is equivalent to that in the standards.
This allowed mapping a cross-correlation 
width to a \vsini\.  The uncertainty from this calibration is of order 1.0 km/s.  The more 
conservative uncertainty  that we quote is estimated by taking the standard deviation in 
the \vsini\ from the cross correlation peak width calibration across multiple orders for both 
epochs of echelle observations.

The ensuing spectroscopic abundance analysis has been performed
differentially with respect to a solar spectrum obtained using the
echelle spectrograph on the CTIO 4-m telescope.  First, we examined
the spectral features for any evidence of the secondary (asymmetry in
cross correlation peaks, double lines).  No such evidence was found.
All lines were measured in both the solar and stellar spectrum and
final abundances were found by subtracting solar abundances from
stellar abundances in a line by line fashion.  Unless otherwise noted,
we quote all abundances in the standard bracket notation where
$[X/Y]=\log\frac{N(X)}{N(Y)}_{\rm
  stellar}-\log\frac{N(X)}{N(Y)}_{\odot}$ and $\log N(H)=12$.

We used 1-D plane-parallel model atmospheres interpolated from the
ATLAS9 grids of Kurucz with an updated version of the LTE Spectral
Synthesis tool MOOG \citep{Sneden73} to derive abundances in a curve
of growth approach from equivalent widths of spectral features.  We
use a high fidelity sample of 52 \ion{Fe}{1} and 6 \ion{Fe}{2}
lines taken from \citet{Bubar10}.

The temperature, surface gravity, microturbulent velocity ($\xi$), and
metallicity have been determined using the approach of
excitation/ionization balance.  The approach is commonly used for
solar-type stars, and there is evidence that it is robust enough for
application to our presumably RS CVn-like system \citep{Morel04}.
Briefly, the (excitation) temperature was found by adjusting the model
atmosphere temperature to remove any correlation in [Fe I/H] versus
excitation potential.  The microturbulence was determined by adjusting
microturbulence to remove any correlations between [Fe I/H] and line
strength (i.e.\ reduced equivalent width).  The surface gravity was
found by forcing the abundance derived from singly ionized Fe lines to
match that derived from neutral lines.  With the availability of 
Hipparcos data, we were also able to obtain parallax-based physical
surface gravities.  We find the ionization-dependent gravity to be
lower than this physical gravity ($\log g_{\rm phys}$=4.15).  This is quantitatively
similar to the differences between ionization balance and physical gravities
observed by e.g. \citet{Fuhrmann98}.  Utilizing this gravity
has a minimal effect on the quoted abundances (typically $\pm$0.04 dex).
Within the uncertainties, this effect is negligible.

The spectroscopic parameters that we derive are $T_{\rm exc} = 5238
\pm 152$ K, $\log g = 3.64 \pm 0.51 $, [Fe/H] = $-0.99 \pm 0.16$ and
$\xi\, = 2.52 \pm 0.44$ \kms. We also derive oxygen abundances from
the near-IR triplet, which is well known to be effected by significant
NLTE effects.  We interpolated within the grids of \cite{Ramirez07}
(using an IDL Spline3 interpolation routine kindly provided by Ivan
Ramirez 2009, personal communication) to perform NLTE corrections and
find a NLTE abundance.  All other abundances have been derived from
equivalent widths using the \emph{abfind} driver of MOOG, with
conservative uncertainties of order $\pm$0.1--0.2 dex.  Results are
given in Table \ref{tab:abun}.  

\begin{deluxetable}{lcr}
\tablewidth{0pt}
\tablecaption{Abundances for V474 Car\label{tab:abun}}
\tablehead{Element & [X/H] & [X/Fe]}
\startdata
Fe    & $-$0.99 & 0.00\\
O     & $-$0.39 & 0.64\\
Na    & $-$1.23 & $-$0.24\\
Mg    & $-$0.72 & 0.27\\
Si    & $-$1.01 & $-$0.02\\
Ca    & $-$0.55 & 0.44\\
\ion{Ti}{1}  & $-$0.44 & 0.55\\
\ion{Ti}{2}  & $-$0.88 & 0.11\\
Mn    & $-$1.45 & $-$0.46\\
Ni    & $-$1.13 & $-$0.14\\
Y     & $-$1.17 & $-$0.18\\
Ba    & $-$1.06 & $-$0.07
\enddata               
\end{deluxetable}

The derived abundances display a clear enhancement in [$\alpha$/Fe]
(for the elements oxygen, magnesium, calcium and titanium
$<[\alpha/{\rm Fe}]>=$0.35) relative to the non-$\alpha$ abundances
(Na, Mn, Ni, Y and Ba, $<[X/{\rm Fe}]>=-$0.17).  The [Si/Fe] differs 
from the trends of the other $\alpha$ elements, 
a result of uncertainties in the temperature and surface gravity.  
Moreover, the
behavior is qualitatively similar to $\alpha$ enhancements observed in a
sample of six halo RS CVn systems \citep{Morel04}.  Considering that the
system is chromospherically active as implied by the H$\alpha$ emission, a
variety of effects are likely at play and could explain the apparent
enhancements.  The presence of NLTE effects resulting in observational
overexcitation/overionization (i.e.\ abundances derived from higher
excitation potential lines and from lines of an ionized species are
larger than their lower excitation potential and neutral counterparts)
cannot be discounted.  While there is no evidence that higher
excitation potential lines or ionized lines yielded increased
abundances in this star this may just be a manifestation of the
excitation/ionization balance approach employed.  Perhaps one of the
most likely physical mechanisms that are impacting the abundance
determinations above is the presence of significant photospheric
inhomogeneities.  Analytic spot models have been utilized by
\citet{Schuler06} to plausibly explain observational overexcitation
and overionization in young open cluster dwarfs.  Attempting to
account for these effects is beyond the scope of this paper and
therefore, we recommend exercising caution in more detailed
interpretations of the abundance results.

However, for our purposes, we can conclude that the observed abundance
signatures (neglecting lithium, which is discussed below) are at least
qualitatively similar to those found by \citet{Morel04},
using a comparable approach.  This provides compelling, though clearly
not definitive, evidence that this is indeed a halo RS CVn system.

We have also determined lithium abundances through spectral synthesis
of the \ion{Li}{1} resonance line, using an updated line list from
\citet{King97} and King (2009, personal communication).  From spectra at
two available epochs, we derived LTE abundances of logN(Li)=2.10 $\pm$
0.20 and 2.20 $\pm$ 0.20, respectively, with the former being from the
higher quality, higher S/N spectra (Figure \ref{lithium}).  From the code of
\citet{Carlsson94}, we determined NLTE corrections and find NLTE Li
abundances of logN(Li) = 2.17 and 2.26, respectively.  This abundance
is not inconsistent with lithium abundance measurements for other
chromospherically active binaries which occupy a similar temperature
range \citep[i.e.\ Fig. 3a of ][]{Barrado97}.  

\begin{figure}
\includegraphics[width=3.5in]{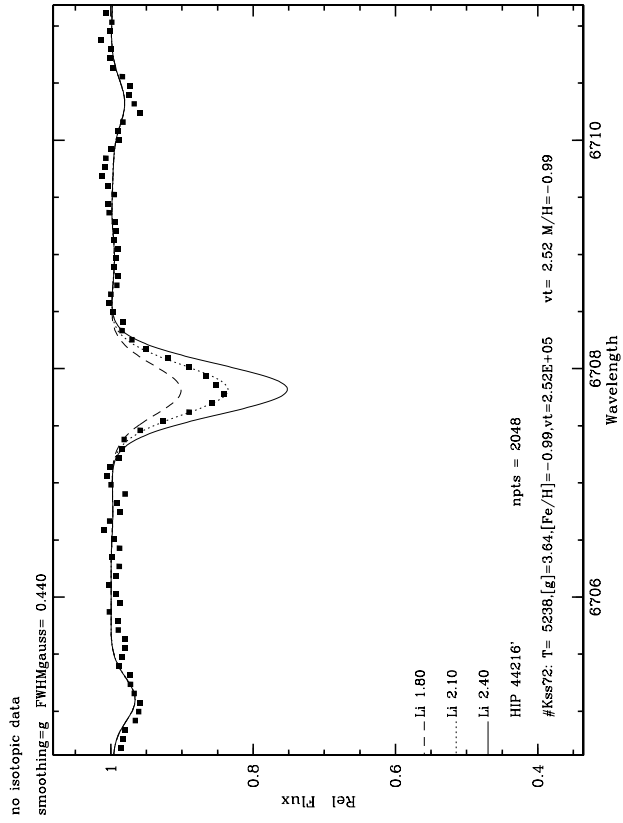}
\caption{Lithium synthesis of the $\lambda$6707 doublet.  A Gaussian
function with a FWHM of 0.44 was used to smooth the synthetic spectrum,
based on measurements of clean, weak lines in the spectrum.  The figure
shows the best fit lithium synthesis of logN(Li)=2.10 from the higher S/N spectrum. 
  \label{figure:lithium}}
\end{figure}

\section{Discussion}

\subsection{Classification}

We can rule out V474 Car being a W UMa contact binary, as its
rotational and orbital periods (10 days) are significantly longer than
the $\sim$0.2--1.2 days typically measured for W UMa systems
\citep{Rucinski97, Maceroni96}. Similarly, we can rule out V474 Car
being a FK Com-type rapidly rotating active G/K giant, as they are
typically single (i.e.\ no velocity variations), and have $v \sin i
\sim 100$ \kms. We can also rule out V474 Car being a BY Dra binary,
as it is clear that the primary is an evolved star. Lastly, we can
dismiss the idea that V474 Car is a high velocity T Tauri star.
Despite its H$\alpha$ emission, strong X-ray emission, lower surface
gravity than dwarf stars, and significant Li absorption, the velocity
of V474 Car is clearly halo-like, and our abundance analysis is
consistent with the star being significantly metal poor ([Fe/H]
$\simeq -1.0$) with $\alpha$ element enhancements reminiscent of other
halo stars. Local T Tauri and other young stellar populations ($< 100$
Myr) typically have metallicities near solar
\citep[e.g.][]{Padgett96}, and velocities within $\sim 20$ \kms\ of
the LSR \citep{Makarov07}.  Given the match between the photometric
and orbital period, we have reason to suspect that most of the
indicators suggestive of youth (H$\alpha$, X-ray emission, starspots)
are due to the system being a tidally-locked binary, rather than
youth.

RS CVn systems are late-type, detached binary stars, with at least one
component being evolved. RS CVns typically have rotation and orbital
periods of $\sim$3--20 days, are photometrically variable, and show
strong chromospheric activity. V474 Car appears to satisfy all of
these characteristics, and hence we classify the star as an RS
CVn-type binary. RS CVns typically have $v \sin i$ values in the tens
of \kms, consistent with what we measure from our spectroscopic
analysis (Table \ref{table:pp}).\footnote{The value of $v \sin i =
  1.0 \pm 1.2$ \kms\ reported by \citet{Torres06} is a typo and should
  be 11.7 \kms\ (C. Melo, personal communication, 2010).}

\subsection{V474 Car: A Halo RS CVn System}

A recent compilation by \citet{Eker08} records $\sim$400 known
chromospherically active binaries (CABs), with 164 explicitly
classified as RS CVn systems. There are indications that V474 Car is
fairly unusual compared to the majority of known RS CVn systems.

We cross-correlated all CABs in \citet{Eker08} with the revised
Hipparcos catalog and the compiled radial velocity catalog of
\citet{Gontcharov06}.  We estimated Galactic Cartesian (UVW)
velocities for all CABs with positive Hipparcos parallaxes and radial
velocities in \citet{Gontcharov06}. For those 251 CABs with velocities
known to better than 50\% accuracy, we find that V474 Car's velocity
(S $= 280 \pm 7$ \kms) exceeds the heliocentric velocity of \emph{all} of these systems.
The highest velocity star from \citeauthor{Eker08}'s compilation is
HIP 81170 with S $= 225 \pm 6$ \kms, the only HIP star in that catalog
with velocity greater than 200 \kms.

\citet{Ottmann97} conducted an X-ray survey of Population II,
metal-poor binaries, and listed a sub-sample of 12 population II
binaries showing significant chromospheric emission (in this case, the
Ca H \& K line). Combining the revised Hipparcos astrometry
\citep{vanLeeuwen07} with the compiled radial velocity catalog of
\citet{Gontcharov06}, we calculated UVW velocities for these 12
emission-line Pop II binaries. We find that only two systems have
heliocentric velocities greater than that of V474 Car: CD$-$48 1741
(HIP 24742) with S $= 377 \pm 28$ \kms, and DR Oct (HD 89499, HIP
49616) with S $= 376 \pm 22$ \kms. CD$-$48 1741 is arguably the most
dynamically similar to V474 Car, with orbital period 7.56 days, [Fe/H] $= -1.40$
\citep{Spite94}, subgiant gravity for the primary \citep[$\log g =$
4.00;][]{Spite94}, and retrograde Galactic orbit ($V = -359 \pm 28$
\kms). Despite its similarity to V474 Car in orbital period and
distance, CD$-$48 1741 was not detected in X-rays in the ROSAT All-Sky
Survey, and its variability is much more subtle (no photometric period
was reported in the Hipparcos catalog or in the AAVSO index).  DR Oct
is a single-lined spectroscopic binary with an evolved G-type primary,
5.57 day orbital period \citep{Ardeberg91}, [Fe/H] $\simeq -2.2$
\citep{Ryan98, Snider01}, enhanced Li compared to halo dwarfs
\citep{Balachandran93}, and it appears to be the most metal-poor star
to have its X-ray spectrum taken \citep{Fleming96}. While DR Oct is
variable, no photometric period has been reported. It appears that
among the \citet{Ottmann97} Pop II binaries, DR Oct may be the most
similar analog to V474 Car, with its retrograde galactic orbit and its
X-ray-luminous, evolved, Li-rich G-type primary.

Starting in 1981, Latham, Carney, and collaborators conducted a long-term spectroscopic
and radial velocity survey of high proper motion
stars, many of which are halo stars. Among the spectroscopic binaries
reported by \citet{Latham02}, they showed that the transition between
apparently tidally-locked ($e \simeq 0$) tight binaries and the
``normal'' wide-separation binaries with a wide range of
eccentricities appears near orbital period $\sim$20 days.  V474 Car's
zero eccentricity orbit is consistent with the $P < 20$ day binary
population observed by \citet{Latham02}. Among the 15 single-lined
spectroscopic binaries from \citet{Latham02} with orbital periods of
$<20$ days, only one has space velocity of $>200$ \kms\, according to
\citet{Carney94}.  That star is G 65-22 (HIP 68527, LHS 2846, Ross
838), the primary of which is an unevolved $\sim$0.6 \msun\, dwarf
with [m/H] $\simeq -1.72$. \citet{Goldberg02} identified 34
double-lined spectroscopic binaries among the Carney-Latham sample.
Of these, only two had orbital periods of $< 20$ days and space
velocity from \citet{Carney94} of $>200$ \kms (G 66-59, G 183-9), and
the primaries for both systems are, again, unevolved metal-poor
$\sim0.6$ \msun\, dwarfs. Hence, the Carney-Latham sample of halo
stars contains a few rare examples of metal-poor, tight binaries that
are {\it dwarfs}, but no evolved specimens like V474 Car.

\citet{Carney05} found that while halo and disk stars have
surprisingly similar binary frequencies and distributions of orbital
parameters, there is a significant difference between the binary
frequency of halo stars in prograde orbits ($\sim$27\%) versus those
in retrograde orbits ($\sim$11\%; like V474 Car).  \citet{Carney05}
proposed that given the difference in binarity may be due to the
retrograde population containing a significant population of stars
from an object (perhaps $\omega$ Cen?) that was captured by the Milky
Way.  If so, V474 Car may be a rare binary member of this purportedly
accreted population.

V474 Car's Galactic orbit is notable for its extreme eccentricity.  In the
Geneva-Copenhagen survey of F/G-type stars in the solar neighborhood
\citep{Holmberg09}, only nine of the 16,682 stars (0.05\%) in the survey
have an eccentricity greater than or equal to that of V474 Car, and
only six of the 16,682 (0.04\%) have a smaller perigalacticon. None of
those stars, however, are detected in X-rays in the ROSAT All-Sky
Survey Bright Source Catalog \citep{Voges99} or Faint Source Catalog
\citep{Voges00}.  All of those stars are metal poor ($-2.30 <$ [Fe/H]
$< -0.88$), as is V474 Car ([Fe/H] $\simeq -1.0$; \S3.5).

\section{Summary}

We have presented a robust analysis of V474 Car, a rare metal poor ([Fe/H]
$\simeq -1.0$) RS CVn binary with halo kinematics. The star was
originally selected in a survey for young, nearby stars due to
enhanced Li and H$\alpha$ emission, however its extreme heliocentric
velocity ($v_r \simeq 264$ \kms) was the first indication that this
was no typical pre-MS star. The star is a single-lined spectroscopic
binary with evolved G-type primary and a probable K/M-type secondary
with orbital period synchronous with the photometrically-constrained
rotational period of the primary (P $\simeq$ 10.3 days). The ASAS
light curve of the primary also manifests a modulation with a $\sim$6
year period and $\sim$0.1 mag amplitude, perhaps due to a long-term
activity cycle.  The star has an eccentricity for its Galactic orbit
surpassed by only $<$0.05\%\, of F/G-type field stars, and its orbit
is mildly retrograde ($V_{\rm helio} = -241$ \kms).  The star has few
peers, with CD$-$48 1741 and DR Oct being perhaps the closest analogs
as rare halo RS CVns. Its X-ray luminosity and hardness of soft X-ray
emission as measured by ROSAT appears similar to that of normal disk
RS CVn stars.  V474 Car may become an important target for studying
the effects of low metallicity on stellar coronae and stellar activity
cycles.

\acknowledgements E.B. would like to thank Jeremy King for useful
discussions regarding the abundance analysis.  Thanks also go to Peter
Teuben for technical assistance with NEMO\null.  We thank Andrej Prsa
and Scott Engle for discussions about the ASAS data, David Cohen
for discussion about the ROSAT data, and the anonymous referee
for suggestions which improved the paper.  E.J. gratefully acknowledges the
support of NSF grant AST-0307830. E.B. and E.M. acknowledge support
from the University of Rochester School of Arts and Sciences.  Stony Brook
University is a member of the SMARTS Consortium, which operates the
SMARTS observatory on Cerro Tololo.

\newpage
       
\begin{deluxetable}{lccccccc}
\setlength{\tabcolsep}{0.03in}
\tablewidth{0pt}
\tablecaption{Atomic Data\label{table:atomic}}
\tablehead{Wavelength & Ion & $\chi$ & log\emph{gf} & EQW$_{\odot}$ & LogN$_{\odot}$ & EQW$_{v474}$ & logN$_{v474}$  \\     
            (\AA)     &     & (eV)   &              & (m\AA)        &                & (m\AA)       & }  
\startdata
 5054.640  &  \ion{Fe}{1}  &  3.64  &  -1.920  &   42.4  &  7.42  &   21.2  &  6.48  \\ 
 5067.160  &  \ion{Fe}{1}  &  4.22  &  -0.970  &   60.9  &  7.37  &   52.2  &  6.68  \\
 5090.780  &  \ion{Fe}{1}  &  4.26  &  -0.400  &   90.9  &  7.34  &   68.1  &  6.36  \\
 5109.660  &  \ion{Fe}{1}  &  4.30  &  -0.980  &   77.9  &  7.74  &   40.9  &  6.61  \\
 5127.370  &  \ion{Fe}{1}  &  0.91  &  -3.310  &   95.3  &  7.29  &   94.0  &  5.96  \\
 5242.490  &  \ion{Fe}{1}  &  3.63  &  -0.970  &   84.0  &  7.27  &   61.1  &  6.16  \\
 5307.360  &  \ion{Fe}{1}  &  1.61  &  -2.990  &   87.7  &  7.46  &   97.8  &  6.46  \\
 5373.700  &  \ion{Fe}{1}  &  4.47  &  -0.760  &   64.4  &  7.43  &   32.7  &  6.42  \\
 5386.340  &  \ion{Fe}{1}  &  4.15  &  -1.670  &   27.9  &  7.33  &    4.7  &  6.04  \\
 5398.280  &  \ion{Fe}{1}  &  4.44  &  -0.630  &   71.5  &  7.39  &   50.3  &  6.53  \\
 5505.880  &  \ion{Fe}{1}  &  4.42  &  -1.300  &   56.6  &  7.78  &   15.8  &  6.52  \\
 5506.780  &  \ion{Fe}{1}  &  0.99  &  -2.800  &  122.1  &  7.26  &  179.9  &  6.68  \\
 5554.880  &  \ion{Fe}{1}  &  4.55  &  -0.440  &   99.6  &  7.71  &   47.8  &  6.41  \\
 5576.090  &  \ion{Fe}{1}  &  3.43  &  -1.000  &  118.7  &  7.61  &   88.3  &  6.30  \\
 5852.220  &  \ion{Fe}{1}  &  4.55  &  -1.230  &   40.4  &  7.51  &   20.1  &  6.69  \\
 5855.080  &  \ion{Fe}{1}  &  4.61  &  -1.480  &   19.6  &  7.34  &    8.2  &  6.57  \\
 5856.090  &  \ion{Fe}{1}  &  4.29  &  -1.330  &   33.8  &  7.23  &   11.3  &  6.23  \\
 5859.590  &  \ion{Fe}{1}  &  4.55  &  -0.300  &   73.6  &  7.17  &   30.5  &  5.99  \\
 5862.360  &  \ion{Fe}{1}  &  4.55  &  -0.060  &   84.4  &  7.11  &   49.5  &  6.04  \\
 5956.690  &  \ion{Fe}{1}  &  0.86  &  -4.600  &   59.1  &  7.62  &   59.2  &  6.68  \\
 6003.010  &  \ion{Fe}{1}  &  3.88  &  -1.120  &   85.9  &  7.62  &   56.2  &  6.48  \\
 6065.480  &  \ion{Fe}{1}  &  2.61  &  -1.530  &  112.0  &  7.28  &  119.5  &  6.32  \\
 6151.620  &  \ion{Fe}{1}  &  2.18  &  -3.300  &   49.3  &  7.44  &   49.3  &  6.70  \\
 6170.500  &  \ion{Fe}{1}  &  4.79  &  -0.440  &   77.2  &  7.56  &   46.7  &  6.62  \\
 6173.340  &  \ion{Fe}{1}  &  2.22  &  -2.880  &   66.1  &  7.39  &   57.0  &  6.42  \\
 6213.430  &  \ion{Fe}{1}  &  2.22  &  -2.480  &   82.1  &  7.32  &   90.1  &  6.44  \\
 6216.360  &  \ion{Fe}{1}  &  4.73  &  -1.420  &   35.7  &  7.76  &   33.8  &  7.34  \\
 6219.280  &  \ion{Fe}{1}  &  2.20  &  -2.430  &   89.0  &  7.38  &   97.6  &  6.46  \\
 6232.640  &  \ion{Fe}{1}  &  3.65  &  -1.220  &   82.8  &  7.43  &   51.7  &  6.25  \\
 6246.320  &  \ion{Fe}{1}  &  3.60  &  -0.730  &  119.3  &  7.45  &  110.9  &  6.47  \\
 6252.550  &  \ion{Fe}{1}  &  2.40  &  -1.690  &  118.8  &  7.32  &  115.4  &  6.18  \\
 6256.360  &  \ion{Fe}{1}  &  2.45  &  -2.410  &   87.6  &  7.57  &   91.9  &  6.64  \\
 6265.130  &  \ion{Fe}{1}  &  2.17  &  -2.550  &   84.1  &  7.37  &   95.4  &  6.52  \\
 6290.970  &  \ion{Fe}{1}  &  4.73  &  -0.780  &   73.3  &  7.79  &   73.3  &  7.24  \\
 6322.690  &  \ion{Fe}{1}  &  2.59  &  -2.430  &   73.4  &  7.45  &   65.6  &  6.48  \\
 6336.820  &  \ion{Fe}{1}  &  3.68  &  -0.910  &  108.4  &  7.54  &   86.2  &  6.41  \\
 6344.150  &  \ion{Fe}{1}  &  2.43  &  -2.920  &   67.1  &  7.65  &   47.3  &  6.55  \\
 6393.610  &  \ion{Fe}{1}  &  2.43  &  -1.570  &  122.7  &  7.27  &  119.2  &  6.13  \\
 6411.650  &  \ion{Fe}{1}  &  3.65  &  -0.590  &  140.0  &  7.55  &  103.2  &  6.27  \\
 6498.940  &  \ion{Fe}{1}  &  0.96  &  -4.700  &   43.5  &  7.47  &   56.1  &  6.82  \\
 6533.940  &  \ion{Fe}{1}  &  4.56  &  -1.380  &   60.2  &  8.01  &    9.2  &  6.45  \\
 6592.910  &  \ion{Fe}{1}  &  2.73  &  -1.470  &  119.5  &  7.38  &  107.0  &  6.18  \\
 6593.870  &  \ion{Fe}{1}  &  2.43  &  -2.420  &   82.5  &  7.43  &   77.8  &  6.43  \\
 6609.110  &  \ion{Fe}{1}  &  2.56  &  -2.690  &   68.3  &  7.55  &   37.2  &  6.30  \\
 6733.150  &  \ion{Fe}{1}  &  4.64  &  -1.580  &   24.0  &  7.55  &   11.2  &  6.82  \\
 6750.150  &  \ion{Fe}{1}  &  2.42  &  -2.620  &   70.4  &  7.37  &   74.2  &  6.56  \\
 6786.860  &  \ion{Fe}{1}  &  4.19  &  -2.070  &   21.5  &  7.55  &    5.2  &  6.47  \\
 7130.920  &  \ion{Fe}{1}  &  4.22  &  -0.790  &   96.2  &  7.67  &   66.8  &  6.59  \\
 7284.840  &  \ion{Fe}{1}  &  4.14  &  -1.750  &   42.3  &  7.61  &   21.7  &  6.75  \\
 7285.270  &  \ion{Fe}{1}  &  4.61  &  -1.700  &   43.0  &  8.02  &   43.2  &  7.60  \\
 7583.790  &  \ion{Fe}{1}  &  3.01  &  -1.890  &   77.7  &  7.29  &   61.1  &  6.27  \\
 7879.750  &  \ion{Fe}{1}  &  5.03  &  -1.650  &   39.4  &  8.27  &    8.9  &  7.14  \\
 6084.110  &  \ion{Fe}{2}  &  3.20  &  -3.800  &   20.0  &  7.47  &    6.0  &  6.40  \\             
 6149.250  &  \ion{Fe}{2}  &  3.89  &  -2.880  &   35.0  &  7.60  &    8.0  &  6.33  \\ 
 6238.390  &  \ion{Fe}{2}  &  3.89  &  -2.750  &   47.0  &  7.75  &   17.0  &  6.57  \\ 
 6247.560  &  \ion{Fe}{2}  &  3.89  &  -2.440  &   53.0  &  7.57  &   21.0  &  6.37  \\ 
 6369.460  &  \ion{Fe}{2}  &  2.89  &  -4.230  &   19.0  &  7.57  &    7.0  &  6.57  \\ 
 6456.380  &  \ion{Fe}{2}  &  3.90  &  -2.070  &   66.0  &  7.49  &   33.0  &  6.27  \\ 
 6300.310  &  \ion{O}{1}   &  0.00  &  -9.72   &    5.6  &  8.68  &   23.9  &  8.22  \\
 7771.940  &  \ion{O}{1}   &  9.15  &   0.37   &   70.0  &  8.72  &   38.3  &  8.51  \\
 7774.170  &  \ion{O}{1}   &  9.15  &   0.22   &   56.7  &  8.67  &   22.6  &  8.29  \\
 7775.390  &  \ion{O}{1}   &  9.15  &   0.00   &   49.1  &  8.74  &   14.5  &  8.24  \\ 
 5688.190  &  \ion{Na}{1}  &  2.11  &  -0.42   &  116.9  &  6.26  &   47.6  &  5.03  \\
 5711.090  &  \ion{Mg}{1}  &  4.35  &  -1.83   &  106.3  &  7.66  &   87.0  &  6.94  \\
 5684.490  &  \ion{Si}{1}  &  4.95  &  -1.55   &   57.1  &  7.41  &   25.5  &  6.55  \\
 5690.430  &  \ion{Si}{1}  &  4.93  &  -1.77   &   56.6  &  7.61  &   14.0  &  6.44  \\
 6155.130  &  \ion{Si}{1}  &  5.62  &  -0.78   &   79.9  &  7.49  &   27.4  &  6.49  \\
 6161.297  &  \ion{Ca}{1}  &  2.52  &  -1.27   &   66.6  &  6.4   &   61.8  &  5.81  \\
 6166.439  &  \ion{Ca}{1}  &  2.52  &  -1.14   &   67.2  &  6.28  &   58.0  &  5.63  \\
 6455.600  &  \ion{Ca}{1}  &  2.52  &  -1.50   &   57.4  &  6.46  &   54.2  &  5.93  \\
 6499.650  &  \ion{Ca}{1}  &  2.52  &  -1.00   &   82.1  &  6.37  &   96.1  &  5.94  \\
 5978.541  &  \ion{Ti}{1}  &  1.87  &  -0.50   &   22.7  &  4.88  &   25.9  &  4.41  \\
 6126.216  &  \ion{Ti}{1}  &  1.07  &  -1.43   &   23.0  &  5.02  &   36.8  &  4.65  \\
 6258.706  &  \ion{Ti}{1}  &  1.46  &  -0.24   &   74.0  &  5.23  &   90.8  &  4.59  \\
 6261.098  &  \ion{Ti}{1}  &  1.43  &  -0.48   &   50.9  &  4.99  &   61.2  &  4.43  \\
 6743.122  &  \ion{Ti}{1}  &  0.90  &  -1.63   &   14.7  &  4.76  &   34.7  &  4.6   \\
 6491.561  &  \ion{Ti}{2}  &  2.06  &  -1.79   &   40.8  &  4.89  &   29.9  &  4.01  \\
 6491.582  &  \ion{Ti}{2}  &  2.06  &  -2.15   &   40.8  &  5.25  &   29.9  &  4.37  \\
 6013.530  &  \ion{Mn}{1}  &  3.07  &  -0.25   &   76.1  &  5.49  &   36.3  &  4.18  \\
 6016.670  &  \ion{Mn}{1}  &  3.08  &  -0.10   &   90.5  &  5.61  &   36.5  &  4.05  \\
 6021.800  &  \ion{Mn}{1}  &  3.08  &   0.03   &   94.2  &  5.53  &   46.3  &  4.06  \\
 6086.280  &  \ion{Ni}{1}  &  4.26  &  -0.51   &   42.3  &  6.26  &   11.5  &  5.11  \\
 6175.370  &  \ion{Ni}{1}  &  4.09  &  -0.53   &   53.8  &  6.34  &   24.5  &  5.32  \\
 6482.810  &  \ion{Ni}{1}  &  1.93  &  -2.97   &   42.8  &  6.44  &   15.1  &  5.16  \\
 6643.640  &  \ion{Ni}{1}  &  1.68  &  -2.01   &   94.2  &  6.23  &   84.8  &  5.02  \\
 6772.320  &  \ion{Ni}{1}  &  3.66  &  -0.98   &   47.4  &  6.23  &   23.4  &  5.26  \\
 5200.413  &  \ion{Y}{2}   &  0.99  &  -0.57   &   46.3  &  2.37  &   31.5  &  1.2   \\
 5853.690  &  \ion{Ba}{2}  &  0.60  &  -1.00   &   63.7  &  2.25  &   78.2  &  1.25  \\
 6141.730  &  \ion{Ba}{2}  &  0.70  &  -0.07   &  110.6  &  2.27  &  130.2  &  1.11  \\
 6496.910  &  \ion{Ba}{2}  &  0.60  &  -0.41   &   94.1  &  2.21  &  122.7  &  1.19 
\enddata               
\tablecomments{Online Table: Line data}
\end{deluxetable}

\end{document}